\documentclass[aps,prl,superscriptaddress, twocolumn, amsmath, amssymb]{revtex4-1}
\usepackage{graphicx}
\usepackage{amsmath,amssymb}
\usepackage{dcolumn}
\usepackage{mathrsfs}
\usepackage{physics}
\usepackage[utf8]{inputenc}
\usepackage{float}
\usepackage{booktabs}
\usepackage{svg}
\usepackage{color}

\begin{document}
\title{A stochastic approach to the quantum noise of a single-emitter nanolaser}
\date{\today}
	
\author{Matias Bundgaard-Nielsen}
\affiliation{Department of Electrical and Photonics Engineering, Technical University of Denmark, Building 343, 2800 Kongens Lyngby, Denmark}
\affiliation{NanoPhoton-Center for Nanophotonics, Technical University of Denmark, Building 343, 2800 Kongens Lyngby, Denmark}
\author{Emil Vosmar Denning}
\affiliation{Department of Electrical and Photonics Engineering, Technical University of Denmark, Building 343, 2800 Kongens Lyngby, Denmark}
\affiliation{NanoPhoton-Center for Nanophotonics, Technical University of Denmark, Building 343, 2800 Kongens Lyngby, Denmark}
\affiliation{Nichtlineare Optik und Quantenelektronik, Institut f\"ur Theoretische Physik, Technische Universit\"at Berlin, Berlin, Germany}
\author{Marco Saldutti}
\affiliation{Department of Electrical and Photonics Engineering, Technical University of Denmark, Building 343, 2800 Kongens Lyngby, Denmark}
\affiliation{NanoPhoton-Center for Nanophotonics, Technical University of Denmark, Building 343, 2800 Kongens Lyngby, Denmark}
\author{Jesper M\o rk}
\affiliation{Department of Electrical and Photonics Engineering, Technical University of Denmark, Building 343, 2800 Kongens Lyngby, Denmark}
\affiliation{NanoPhoton-Center for Nanophotonics, Technical University of Denmark, Building 343, 2800 Kongens Lyngby, Denmark}

\begin{abstract}
It is shown that the quantum intensity noise of a single-emitter nanolaser can be accurately computed by adopting a stochastic interpretation of the standard rate equation model. The only assumption made is that the emitter excitation and photon number are stochastic variables with integer values. This extends the validity of rate equations beyond the mean-field limit and avoids using the standard Langevin approach, which is shown to fail for few emitters. The model is validated by comparison to full quantum simulations of the relative intensity noise and second-order intensity correlation function, $g^{(2)}(0)$. Surprisingly, even when the full quantum model displays vacuum Rabi oscillations, which are not accounted for by rate equations, the intensity quantum noise is correctly predicted by the stochastic approach. Adopting a simple discretization of the emitter and photon populations, thus, goes a long way in describing quantum noise in lasers. Besides providing a versatile and easy-to-use tool for modeling emerging nanolasers, these results provide insight into the fundamental nature of quantum noise in lasers. 
\end{abstract}
\maketitle

The ability of a laser to generate a coherent optical signal with ultra-low noise is key to a wide range of applications, including the internet \cite{Ning2019SemiconductorReview}, sensors \cite{Ma2019ApplicationsNanolasers}, as well as fundamental tests of physics \cite{Abbott2016ObservationMerger}. Recent advances in nanotechnology have enabled the realization of a new generation of microscopic lasers, for instance, based on semiconductor quantum dots in photonic crystals \cite{Notomi2010ManipulatingCrystals,Ota2017ThresholdlessNanolaser,Saldutti2021ModalLasers}, opening new possibilities in, e.g., on-chip communications \cite{J.Shapiro1979OpticalPerformance,Saleh1992InformationLight,Notomi2014TowardChip,Miller2017AttojouleCommunications} and quantum technology \cite{Weedbrook2012GaussianInformation}. However, as the laser shrinks into the microscopic regime, the power decreases, and the intrinsic quantum noise of the laser may lead to unacceptable bit-error rates in optical links \cite{Mork2020SqueezingConfinement}. The noise of lasers is a rich and complex field which is still developing \cite{Kozlovskii2014Super-poissonianThreshold,Kreinberg2017EmissionCoupling, Gies2017StrongLasing, Mork2018RateEmitters, Andre2020, Takemura2021Low-Analogy, Protsenko2021QuantumNanolasers, Dimopoulos2022Electrically-DrivenThreshold, Yacomotti2023QuantumLimit, Carroll2021ThermalDevices, Vyshnevyy2022CommentDevices,Carroll2022CarrollReply:,Wang2020Superthermal-lightLasers}. For few emitters, it is, in principle, possible to perform full-scale quantum simulations of the noise properties \cite{Mu1992One-atomLasers,Loffler1997SpectralLaser,DelValle2011RegimesExcitation,Poshakinskiy2014Time-dependentMode,Clemens2004SpectraLasers}. However, such simulations are numerically demanding for more than a few emitters. Instead, the use of rate equations has proven extremely successful in realizing the advanced semiconductor laser of today \cite{Coldren1997}. However, as we shall show, rate equations, even with the addition of stochastic Langevin noise terms \cite{Roy-Choudhury2010QuantumDiodes,Moelbjerg2013DynamicalEmitters}, do not correctly account for the quantum noise of few-emitter lasers. 
 
\begin{figure}
    \centering
    \includegraphics[width=\linewidth]{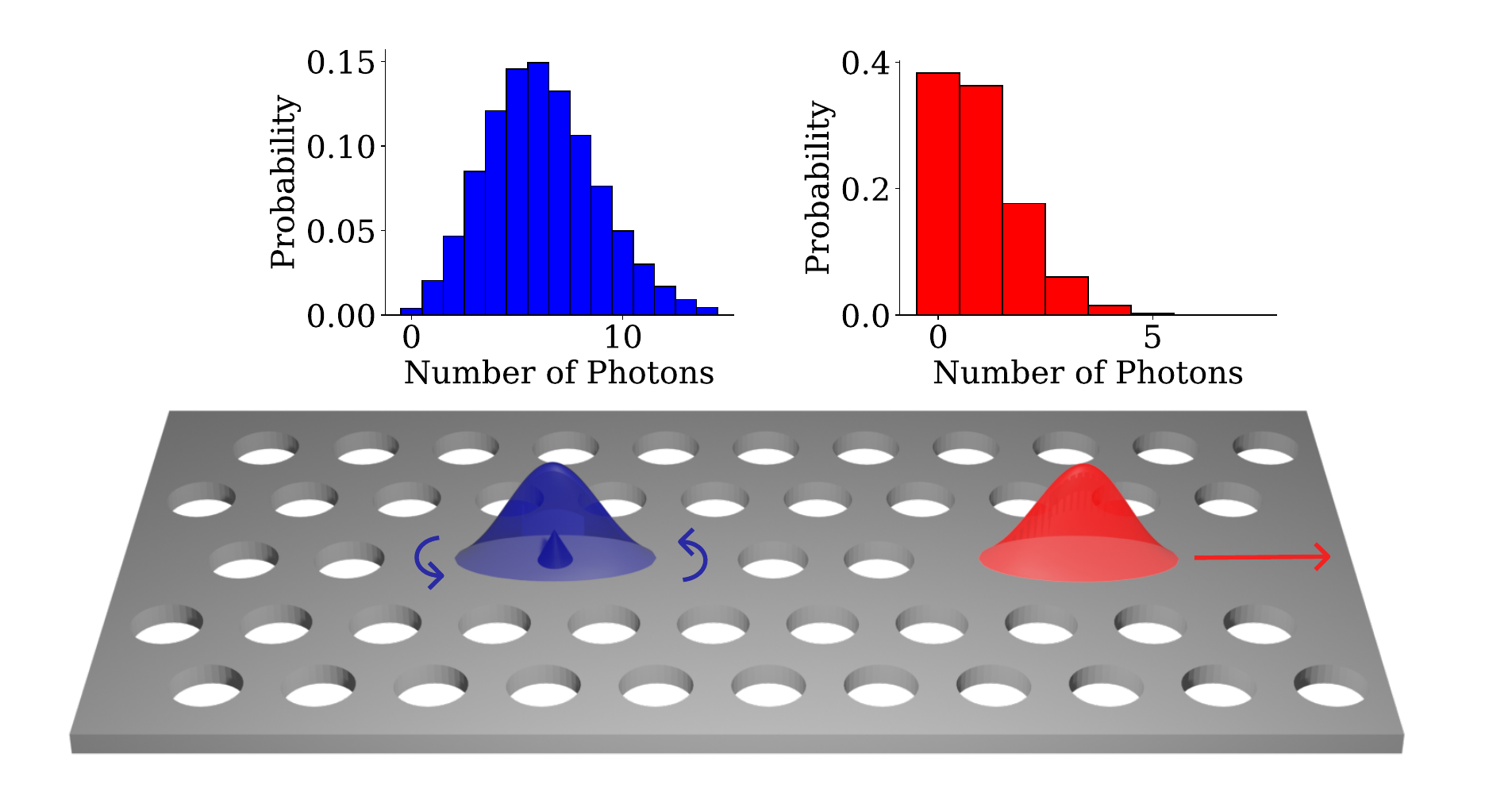}
    \caption{Schematic of a photonic crystal cavity laser containing a single quantum dot. Examples of photon distributions inside and outside the cavity, calculated with the stochastic approach, are shown above threshold for the same parameters as in Figs.\ \ref{Fig:rin_spectrum} and \ref{Fig:g2_tau}.  }
    \label{Fig:sketch}
\end{figure}

In this paper, we consider the ultimate limit of a nanolaser, where the gain medium is a single-emitter, e.g., a quantum dot in a photonic crystal cavity as studied in \cite{Nomura2010LaserSystem,Strauf2011SingleNanolaser}, and illustrated in Fig.\ \ref{Fig:sketch}. We show that the quantum noise of such nanolasers can be quantitatively described by adopting a simple stochastic interpretation of conventional rate equations in terms of discrete rather than continuous variables. The appearance of sub-Poissonian statistics below threshold is thus predicted by our approach, while standard Langevin approaches are shown to fail in this regime. The stochastic approach is thereby shown to be a more correct way of adding quantum noise to rate equations. Furthermore, the stochastic approach is easily extended to more complex systems and scenarios, such as large-signal temporal modulation, where analytical small-signal results are inapplicable.

Our finding enables a new approach toward the quantum noise of nanolasers. Not only does it provide an intuitive and simple simulation tool, but it also offers new insights into the origin of quantum noise in lasers.

\label{sec:intro}

In second quantization, the single-emitter laser is described by a master equation (ME) of the form \cite{Loffler1997SpectralLaser,Mu1992One-atomLasers}:

\begin{equation}
    \frac{\partial \rho}{\partial t} = -\frac{i}{\hbar} \comm{H}{\rho} + \kappa \mathcal{D}_{a}[\rho] + \gamma_D \mathcal{D}_{\sigma^\dagger \sigma}[\rho] +\gamma_A \mathcal{D}_{\sigma}[\rho] + P \mathcal{D}_{\sigma^\dagger}[\rho] 
    \label{eq:master}
\end{equation}
where $H = -\hbar \Delta a^\dagger a + \hbar g (\sigma^\dagger a + a^\dagger \sigma)$ is the Jaynes-Cummings Hamiltonian with $g=\sqrt{d^2\omega_{eg}/\left(2\hbar\epsilon_0\epsilon V\right)}$ being the light-matter coupling \cite{Denning2020OpticalInteractions}. Here $d$ is the emitter dipole moment, $\epsilon$ is the dielectric constant of the background material, $V$ is the cavity mode volume, $\sigma = \ket{g}\bra{e}$ is the atomic transition operator, $a$ is the cavity mode annihilation operator, $\Delta = \omega_{eg} - \omega_c$ is the detuning between the electronic transition $\omega_{eg}$ and the cavity frequency $\omega_c$, and $\mathcal{D}_{A} [\cdot] =  \frac{1}{2}(2A (\cdot) A^\dagger-(\cdot) A^\dagger A - A^\dagger A (\cdot))$ is the Lindblad operator. The various Lindblad terms describe dissipative processes relevant to single-emitter lasers; $\kappa$ is the cavity decay rate, $\gamma_D$ is the pure dephasing rate, arising from, e.g., phonons in quantum dot emitters \cite{Strauf2011SingleNanolaser}, $\gamma_A$ is the non-radiative decay and/or decay into non-lasing modes, and $P$ is the pump rate of the emitter, modeled as incoherent pumping \cite{Mu1992One-atomLasers, Moelbjerg2013DynamicalEmitters}. The ME is numerically implemented by using QuTiP \cite{Johansson2012QuTiP:Systems, Johansson2013QuTiPSystems}.

Under the assumption of a large dephasing rate, such that the polarization can be eliminated \cite{Strauf2011SingleNanolaser, Lorke2013TheoryTheory, Moelbjerg2013DynamicalEmitters}, a rate equation can be derived from the ME in eq.\ \eqref{eq:master}. With $n_a = \expval{a^\dagger a}$ and $n_e = \expval{\sigma^\dagger \sigma}$ and making a mean-field approximation we get:

\begin{align}
      &\frac{d n_a}{dt} = \gamma_r (2n_e-1)n_a + \gamma_r n_e- \kappa n_a \label{eq:na} \\
      &\frac{d n_e}{dt} = P (1-n_e) - \gamma_r (2n_e-1)n_a -\gamma_r n_e - \gamma_A n_e \label{eq:ne}
\end{align}
with an emitter-cavity coupling rate given by $\gamma_r= 4g^2/(P + \kappa + \gamma_D + \gamma_A)$. Since the polarization was adiabatically eliminated, this model does not display Rabi oscillations. Furthermore, the equations only govern the average emitter excitation and number of cavity photons and do not include quantum noise. Conventionally, quantum noise is accounted for by adding random Langevin forces to the RHS. of eqs.\ \eqref{eq:na} and  \eqref{eq:ne} \cite{Coldren1997,Hofmann2000ThermalThreshold,Hofmann2000CoexistenceDiodes}. As we shall see, however, this leads to incorrect results for the intensity correlation, $g^{(2)}(0)$, which is an essential parameter for identifying the regime of lasing \cite{Kreinberg2017EmissionCoupling,Takemura2021Low-Analogy}.

Another approach to include noise in the rate equations is to interpret Eqs.\ \eqref{eq:na} and \eqref{eq:ne} as a stochastic process for integer-valued variables, $n_e$ and $n_a$ \cite{Puccioni2015StochasticLasing,Mork2018RateEmitters}. Thus far, the stochastic approach (StA) has been used only for mesoscopic lasers comprising ten or more emitters \cite{Mork2018RateEmitters}, and it was not expected to be valid for fewer emitters. In essence, it replaces the rates in eqs.\ \eqref{eq:na}-\eqref{eq:ne} by Poisson processes, thus attributing all the quantum noise of the laser to the discrete nature of photons and emitter excitation. Notably, this approach does not require the calculation of diffusion coefficients for Langevin forces, nor does it assume small perturbations around a steady state. Here, to numerically solve the stochastic equation, we use Gillespies first-reaction method \cite{Gillespie1976AReactions,Pahle2008BiochemicalApproaches,Andre2020}.

\begin{figure}[H]
    \centering
    \includegraphics[width = \linewidth]{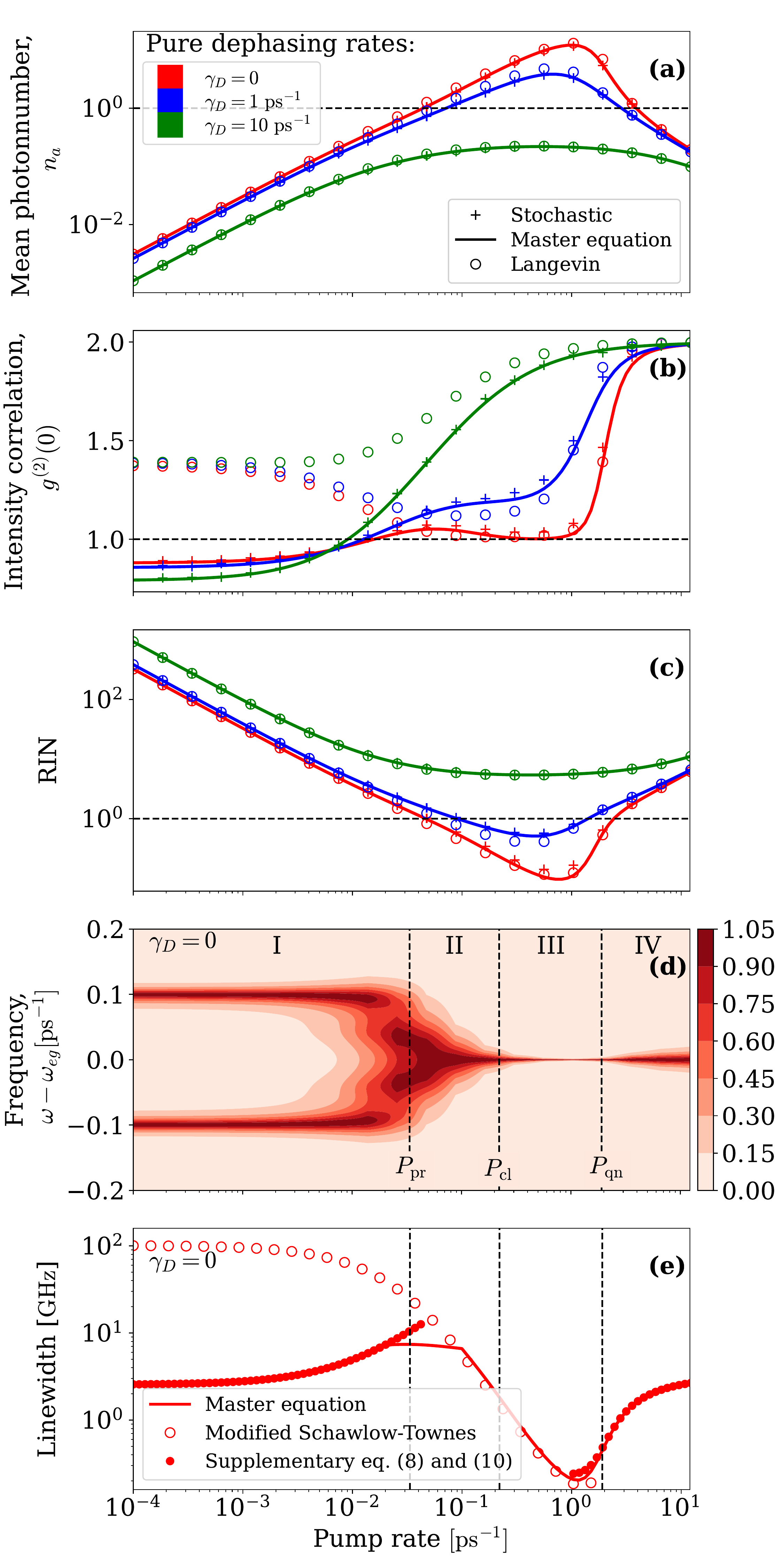}
    \caption{(a) Cavity population, (b) second-order correlation function $g^{(2)}(0)$, (c) RIN, (d) emission spectrum, and (e) linewidth vs. pump-rate, for three different pure dephasing rates: $\gamma_D = 0$ (red), $\gamma_D = 1 \ \mathrm{ps}^{-1}$ (blue), and $\gamma_D = 10 \ \mathrm{ps}^{-1}$ (green). The spectrum, calculated using the ME, is only shown for $\gamma_D = 0$, normalized to 1 for each pump rate. The characteristic pump rates, $P_\mathrm{re}$, $P_\mathrm{cl}$, and $P_\mathrm{qu}$ separate the laser into four qualitatively different regimes I-IV.}
    \label{Fig:pop_emission}
\end{figure}

We choose parameters compatible with a single quantum dot in a photonic crystal cavity with a light-matter coupling of $g=0.1\ \mathrm{ps}^{-1}$ \cite{Nomura2010LaserSystem} and a cavity decay rate $\kappa = 0.02  \ \mathrm{ps}^{-1}$ \cite{Dimopoulos2022Electrically-DrivenThreshold}. Furthermore, we use $\gamma_A = 0.012 \ \mathrm{ps}^{-1}$ and study three different pure dephasing rates $\gamma_D=0,1,10 \ \mathrm{ps}^{-1}$. Ignoring pump broadening, this gives $\beta$-factors of $\beta = \gamma_{r}/(\gamma_{r}+\gamma_A) = 0.999,0.764,0.249$, thus placing the laser in the high-$\beta$ or "thresholdless" regime. It is worth noting that recent advances in dielectric nanocavities with deep subwavelength confinement \cite{Hu2016DesignConcentration,Choi2017Self-SimilarNonlinearities,Wang2018MaximizingCavities, Albrechtsen2022Nanometer-scaleCavities} enable even larger values of $g$.   
Fig.\ \ref{Fig:pop_emission} shows the results obtained from the ME and the StA. The mean photon number $n_a$, the intensity correlation function $g^{(2)}(0)$, Relative Intensity Noise (RIN), emission spectrum, and linewidth are depicted. See Supplementary Material for details.

Fig.\ \ref{Fig:pop_emission} (a,b,c) demonstrates excellent agreement between the StA and the ME for the mean photon number, intensity correlation, and RIN. This is the case for all dephasing values and pump rates. The result of adding Langevin noise terms to eqs.\ \eqref{eq:na} and \eqref{eq:ne}, see \cite{Mork2018RateEmitters} for details, is also shown. It is clear that the Langevin approach (LA) captures the mean photon number and RIN relatively well, while there is a large deviation in the intensity correlation. This shows a fundamental problem of the LA for few emitters, since $g^{(2)}(0)$ is a key measure of the statistics of the light \cite{Moelbjerg2013DynamicalEmitters,Kreinberg2017EmissionCoupling,Gies2017StrongLasing}. Below threshold, the light is antibunched in the single-emitter case, which is correctly identified by $g^{(2)}(0)<1$ for the ME and StA, while the LA predicts super-Poissonian statistics, $g^{(2)}(0)>1$.

The shortcoming of the LA stems from its inability to correctly estimate second-order quantities such as $\expval{n_a^2}$, which $g^{(2)}(0)$ depends very sensitively on. The LA assumes perturbations around the mean value to be small. This is clearly not true for a single emitter below threshold, where the photon population is close to zero. A single spontaneous emission event will here lead to a temporal fluctuation much larger than the mean value itself. On the other hand, the StA does not assume small perturbations and correctly predicts $g^{(2)}(0)$. See Supplementary Material for a derivation of the LA perturbation strength and further details.

Fig.\ \ref{Fig:pop_emission}(d) shows the emission spectrum for the case of $\gamma_D = 0$. Note that the StA, which in its present form does not contain information about the phase, cannot predict the emission spectrum. We observe two spectral peaks for low pump values that reflect Rabi oscillations and correspondingly have a splitting of $2g=0.2 \ \mathrm{ps}^{-1}$. As the pump rate is increased, we see the transition to lasing as the Rabi peaks coalesce into a single peak at the cavity frequency, whose linewidth narrows significantly. This is seen from Fig.\ \ref{Fig:pop_emission}(e), which shows the corresponding linewidth $\Delta \nu$ (FWHM) calculated from the Liouville gap (the smallest real eigenvalue in the system) \cite{Takemura2021Low-Analogy}. 

In contrast to macroscopic lasers, characterized by $\beta << 1$, the transition to lasing in nanolasers with $\beta$ of order unity does not show a clear phase transition \cite{Rice1994PhotonAnalogy}, giving rise to vivid discussions of the proper definition of threshold \cite{Chow2014EmissionLasing,Takemura2019LasingLasers,Lippi2022PhaseNanolasers,Yacomotti2023QuantumLimit}. This highlights the ambiguity in defining the threshold for nanolasers, in contrast to the case of macroscopic lasers. In Fig.\ \ref{Fig:pop_emission}(d,e), vertical lines show the predictions of two threshold definitions, $P_{\mathrm{pr}}$ and $P_{\mathrm{cl}}$, to be further discussed below. In both cases, the number of carriers at lasing threshold, $n_{e,th}$, is defined by the balance between gain and cavity loss $\gamma_r(2 n_{e,th} -1) = \gamma_c$. From here, the classical approach \cite{Coldren1997} is to compute the corresponding pump rate by assuming that the photon population below threshold is zero. This procedure leads to the following expression, where we have ignored pump broadening \cite{marco_unpublished}
\begin{equation}
    P_{\mathrm{cl}} = \left( \frac{2}{1-1/(2 \xi)} \right)\frac{1}{2}\frac{\gamma_c}{\beta}(1+2\xi)
    \label{eq:th_photon_standard}
\end{equation}
with $\xi = \gamma_r/(2\gamma_c)$. However, for a near-unity $\beta$-factor, the number of photons below transparency will be non-negligible \cite{Yamamoto1991Bjork_1991_Jpn._J._Appl._Phys._30_L2039} and a generated photon has a significant chance of being re-absorbed rather than escaping the cavity. This cycle of spontaneous emission into the lasing mode and stimulated re-absorption, i.e., photon recycling \cite{Yamamoto1991Bjork_1991_Jpn._J._Appl._Phys._30_L2039,marco_unpublished}, may effectively increase the carrier lifetime and lower the pump rate required to reach the lasing threshold. By including the effect of photon recycling, one arrives at the following expression \cite{marco_unpublished}:      
\begin{equation}
    P_{\mathrm{pr}} = \left( \frac{2}{1-1/(2 \xi)} \right)\frac{1}{2}\frac{\gamma_c}{\beta_\mathrm{eff}}, \ \ \ \beta_\mathrm{eff} = \frac{\beta}{1+2\xi (1-\beta)} \label{eq:th_photon_recycling}
\end{equation}

From Fig. \ref{Fig:pop_emission}, it is seen that $P_{\mathrm{pr}}$ marks the pump value at which  $g^{(2)}(0)$ approaches one from below, the Rabi peaks coalesce, and the linewidth starts narrowing following the collapse. At the larger pump value of $P_{\mathrm{cl}}$, the linewidth has already reduced significantly, and one enters a regime with $g^{(2)}(0) \approx 1$, independently of the pump value. However, when various light-matter coupling rates are considered (not shown here), it is clear \cite{marco_unpublished} that $P_\mathrm{cl}$ does not mark this point consistently, whereas $P_\mathrm{pr}$ always corresponds to the onset of lasing.

 At a larger critical pump value, denoted as the quenching threshold and indicated by $P_\mathrm{qn}$, the linewidth starts to rebroaden, and $g^{(2)}(0)$ quickly approaches the thermal value of 2. This quenching behavior at high pump values is in agreement with previous work \cite{Loffler1997SpectralLaser,Mu1992One-atomLasers,Moelbjerg2013DynamicalEmitters} and occurs because pump-induced dephasing dominates the emitter broadening. See Supplementary Material for details on $P_\mathrm{qn}$ and also analytical expressions for the linewidth, which are compared to the simulations of the ME in Fig.\ \ref{Fig:pop_emission}(e).

To further characterize the quantum noise, we consider the frequency dependence of the RIN spectrum.  The spectrum of the outcoupled signal is the experimentally relevant observable and differs qualitatively from the intra-cavity spectrum due to the non-trivial action of the outcoupling mirror \cite{Yamamoto1986AmplitudeOscillator,Hofmann2000CoexistenceDiodes}. This is illustrated in Fig.\ \ref{Fig:sketch}, where the photon distribution changes drastically outside the cavity. We calculate the outcoupled noise spectrum by simulating the detection of photons outside the cavity for a finite time $T$. In the ME, this is done using normally ordered photodetection theory \cite{Carmichael1987SpectrumTreatment,Mu1992One-atomLasers,Marte1989LasersPump,Ritsch1990QuantumAbsorbers}, and for the stochastic approach, we track all outcoupling events \cite{Mork2020SqueezingConfinement}; see Supplementary Material for details. In the calculations, we choose a detector integration time small enough to capture all features in the outcoupled spectrum. Empirically, the inverse emission rate $\gamma_r^{-1}$ is a good choice. 
\begin{figure}  
    \centering
    \includegraphics[width=\linewidth]{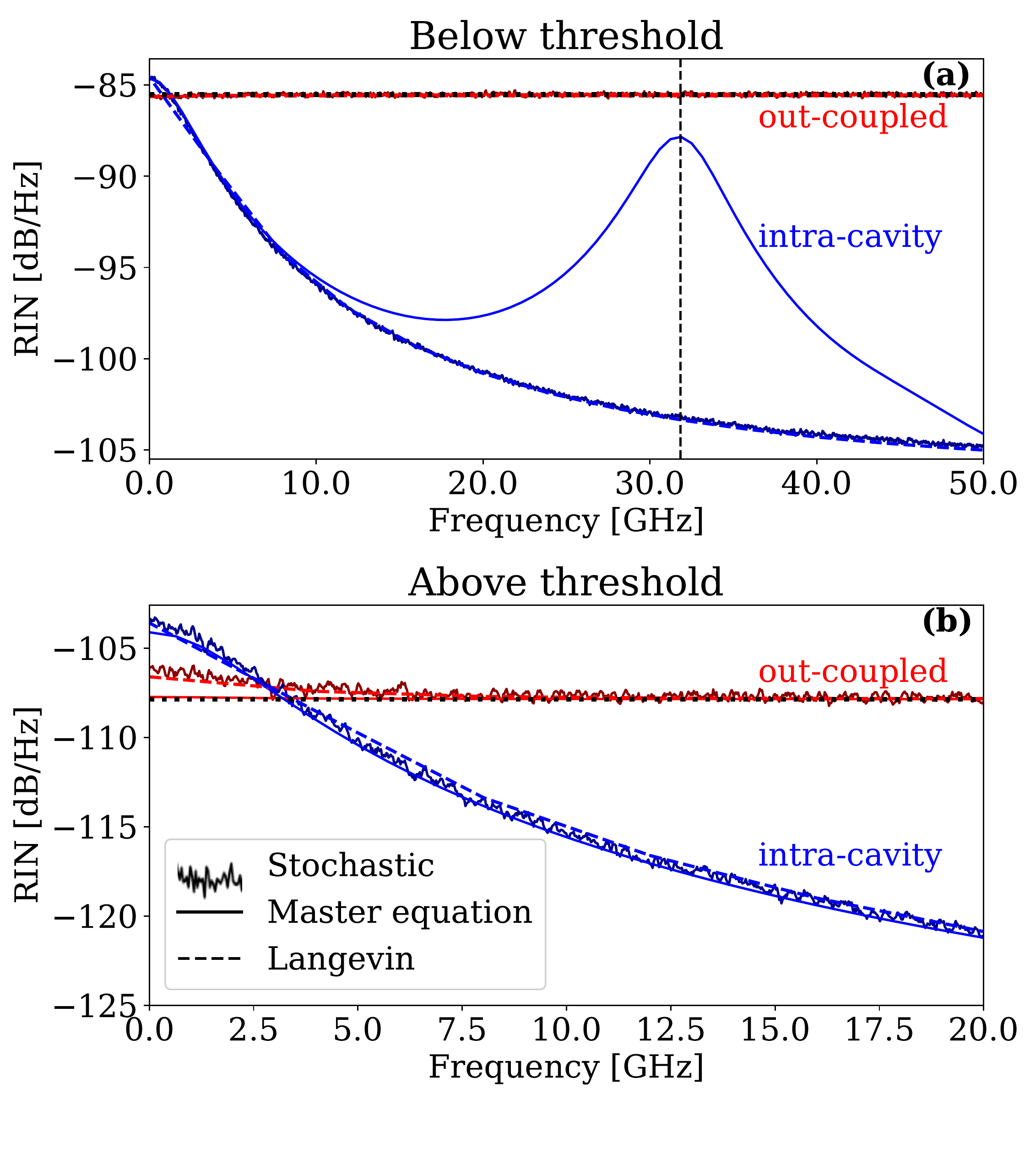}
    \caption{RIN spectra for intra-cavity and outcoupled photons (a) below and (b) above threshold. The parameters are the same as in Fig.\ \ref{Fig:pop_emission} with pump-rates $P = 0.0012 \ \mathrm{ps}^{-1}$ and $P = 0.3023 \ \mathrm{ps}^{-1}$ and detector integration times $T=\gamma_r^{-1} = 0.83 \ \mathrm{ps}^{-1} \ \mathrm{and} \  8.36 \ \mathrm{ps}^{-1}$, respectively below and above threshold. The vertical dashed line shows the Rabi frequency at $f = 2g/2 \pi=31.8 \mathrm{GHz}$, and the horizontal dots mark the standard quantum limit $\mathrm{SN} = 2/(n_a \kappa)$. } 
    \label{Fig:rin_spectrum}
\end{figure}

Fig.\ \ref{Fig:rin_spectrum} shows RIN spectra for the intra-cavity and outcoupled photons for two pump values: one below threshold and one above. We also show results based on the analytic LA introduced in ref. \cite{Coldren1997} and adopted to the nanolaser rate eqs.\ in ref. \cite{Mork2020SqueezingConfinement}. Below threshold, Rabi oscillations manifest themselves in the intra-cavity RIN spectrum calculated by the ME as a peak around $2 \pi f = 2g$. These oscillations arise due to the dynamics of the atomic polarization, which neither the LA nor the StA can capture, due to the adiabatic elimination imposed in the rate equations \cite{Andre2020,Mork2020SqueezingConfinement}. The outcoupled RIN shows that the partition noise at the cavity mirrors \cite{Coldren1997} dominates, and the spectrum is accurately given by the standard quantum limit: $\mathrm{SN} = 2/(n_a \kappa)$, without any features of Rabi oscillations. 

Above threshold, Rabi oscillations do not manifest themselves in the RIN spectrum, and all three approaches agree. At large frequencies shot noise again dominates the outcoupled RIN spectrum. Note, however, that the outcoupled RIN spectrum is not simply given by the intra-cavity RIN spectrum with the added shot noise. Below $5 \ \mathrm{GHz}$, the outcoupled RIN is thus smaller than the intra-cavity RIN \cite{Coldren1997,Mork2020SqueezingConfinement}. While outcoupling at the laser mirror introduces quantization (shot) noise of the outcoupled photons, the noise is reduced at low frequencies due to anti-correlation effects \cite{Coldren1997,Mork2020SqueezingConfinement}.  The partition noise at the cavity mirrors can thus lower the noise in the low-frequency range. 

It should be emphasized that the StA still inherits the adiabatic elimination of the medium polarization from the rate equations. Therefore, the inability of the StA to capture Rabi oscillations was expected. The StA should, therefore, not be seen as a complete alternative to master equations but rather as a more consistent way of adding quantum noise to rate equations. This is especially true in the limit of few emitters, where the assumptions implicit in Langevin equations of small-signal white-noise perturbations around the steady-state are not valid.

\begin{figure}
    \centering
    \includegraphics[width=\linewidth]{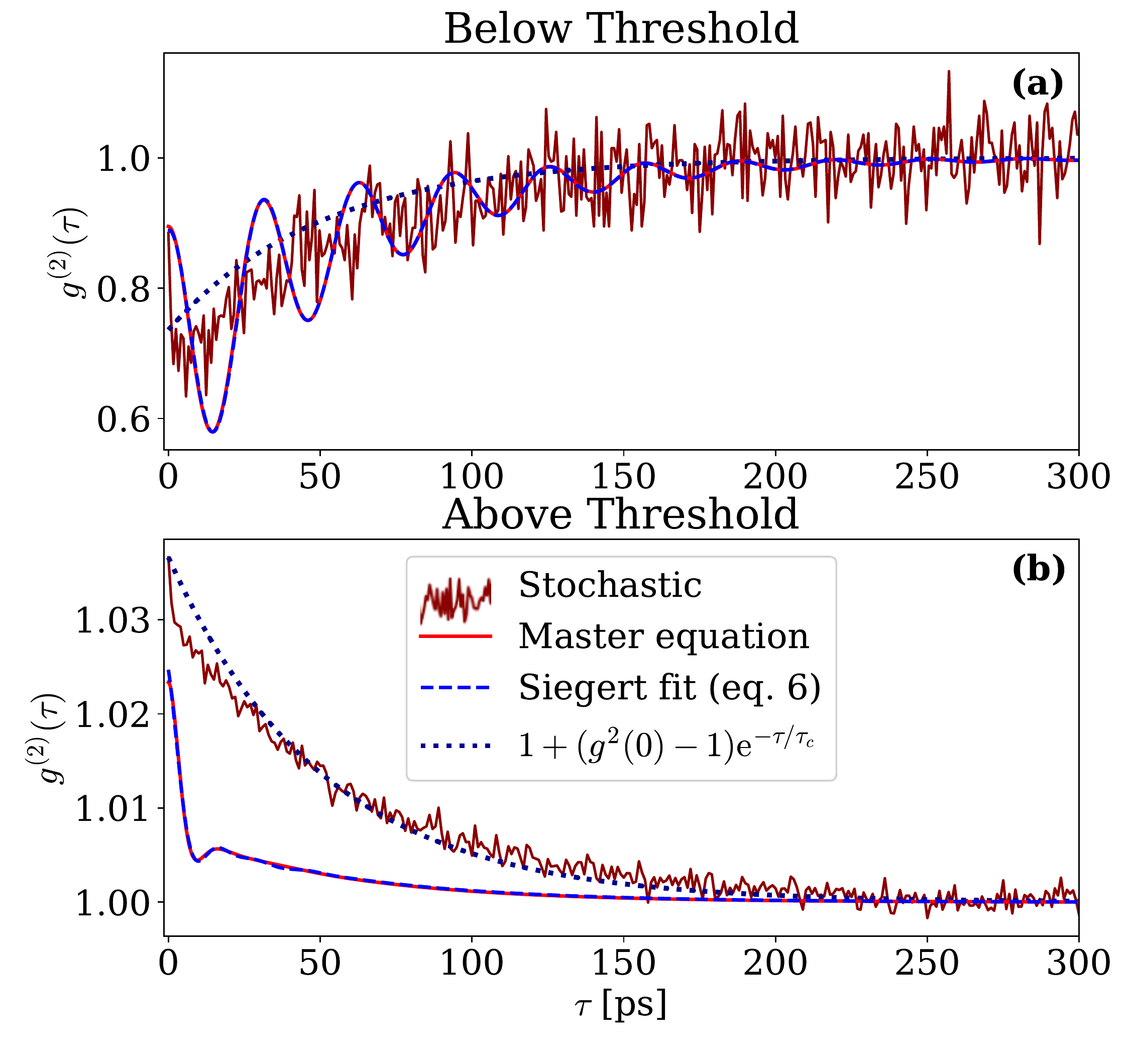}
    \caption{Computed correlation function $g^{(2)}(\tau)$ (a) below and (b) above the laser threshold for the same parameters as in Fig.\ \ref{Fig:pop_emission} and \ref{Fig:rin_spectrum}. A fit to the master equation with the expression \eqref{eq:siegert} is also shown (see Supplementary Material), as well as a monotonically decaying exponential with coherence lifetime $\tau_c^{-1} = \kappa-P/4g^2$.}
    \label{Fig:g2_tau}
\end{figure}

The limitations of the StA are more clearly seen when considering the time-dependency of the intra-cavity intensity correlation function, $g^{2}(\tau)$ (see Supplementary Material). The results can be seen in Fig.\ \ref{Fig:g2_tau}, where we see deviations between the StA and the ME below and above threshold. The deviation below threshold clearly arises from Rabi oscillations, which, as mentioned, cannot be captured by the StA. Above threshold, the absolute deviation is quite small (within 1-2\%), but the qualitative behavior is significantly different. Although not visible in the spectrum, transitions in the Jaynes-Cummings Ladder \cite{Gerry2004IntroductoryOptics} still affect the two-time correlation function. We show this by fitting an expression similar to the Siegert relation found in ref.~\cite{Drechsler2022RevisitingNanolasers} to our results: 
\begin{equation}
    g^{(2)}(\tau) = 1 + (g^{2}(0)-1)\mathrm{e}^{-\tau/\tau_c} + \delta g^{(2)}(\tau) \label{eq:siegert}
\end{equation}
with $\tau_c^{-1} = \kappa-P/4g^2$ \cite{Poshakinskiy2014Time-dependentMode}. See Supplementary Material for details on $\delta g^{(2)}(\tau)$. The correction $\delta g^{(2)}(\tau)$ is necessary when emitter-photon correlations are present \cite{Drechsler2022RevisitingNanolasers}, and when omitted, we see that we recover $g^{2}(\tau)$ as obtained from the StA. The StA thus quantitatively accounts for the zero-time delay intensity correlation of the laser light. This is the critical parameter characterizing the quantum statistics of any light source, whereas time-dependent quantum correlations, $\delta g^{(2)}(\tau)$ are predicted with less accuracy.

In conclusion, we have shown that a simple stochastic interpretation \cite{Andre2020} of standard rate equations accurately accounts for the intensity quantum noise of a single-emitter laser. This implies that the intensity
quantum noise of a laser originates solely from the discrete
nature of photon and emitter excitations. In contrast, the conventional Langevin approach \cite{Coldren1997} does not correctly predict the quantum statistics of light in this regime. We also analyzed the single-emitter lasing transition in detail and introduced a new threshold definition that reliably predicts the onset of lasing. The stochastic approach can easily be extended to multiple emitters \cite{Mork2018RateEmitters,marco_unpublished}, where quantum master equations become too numerically demanding, and to large-signal temporal modulation, where Langevin approaches become inapplicable. Our findings may, therefore, facilitate the analysis and design of a new generation of nanolasers while also allowing for a better understanding of quantum noise.

\begin{acknowledgments}
\section*{ACKNOWLEDGEMENTS}
This work was supported by the Danish National Research Foundation through NanoPhoton - Center for Nanophotonics, Grant No. DNRF147.
EVD acknowledges support from Independent Research Fund Denmark through  an International Postdoc fellowship, grant no. 0164-00014B.
\end{acknowledgments}

\end{document}